\documentclass[conference]{IEEEtran}
\IEEEoverridecommandlockouts
\usepackage{cite}
\usepackage{amsmath,amssymb,amsfonts}
\usepackage{algorithmic}
\usepackage{graphicx}
\usepackage{textcomp}
\usepackage{xcolor}
\def\BibTeX{{\rm B\kern-.05em{\sc i\kern-.025em b}\kern-.08em
    T\kern-.1667em\lower.7ex\hbox{E}\kern-.125emX}}
\begin{document}

\title{Securing Smart Homes via Software-Defined Networking and Low-Cost Traffic Classification}
\author{\IEEEauthorblockN{Holden Gordon, Christopher Batula, Bhagyashri Tushir, Behnam Dezfouli, Yuhong Liu }
\IEEEauthorblockA{Internet of Things Research Lab, Computer Science and Engineering, Santa Clara University, USA\\
\{hgordon, cbatula, btushir, bdezfouli, yhliu\}@scu.edu
}
 }
 
\maketitle

\begin{abstract}
IoT devices have become popular targets for various network attacks due to their lack of industry-wide security standards.
In this work, we focus on the classification of smart home IoT devices and defending them against Distributed Denial of Service (DDoS) attacks.
The proposed framework protects smart homes by using VLAN-based network isolation.
This architecture includes two VLANs: one with non-verified devices and the other with verified devices, both of which are managed by a SDN controller.
Lightweight, stateless flow-based features, including ICMP, TCP and UDP protocol percentage, packet count and size, and IP diversity ratio, are proposed for efficient feature collection. 
Further analysis is performed to minimize training data to run on resource-constrained edge devices in smart home networks. 
Three popular machine learning models, including K-Nearest-Neighbors, Random Forest, and Support Vector Machines, are used to classify IoT devices and detect different DDoS attacks based on TCP-SYN, UDP, and ICMP. 
The system's effectiveness and efficiency are evaluated by emulating a network consisting of an Open vSwitch, Faucet SDN controller, and flow traces of several IoT devices from two different testbeds. 
The proposed framework achieves an average accuracy of 97\%in device classification and 98\% in DDoS detection with average latency of 1.18 milliseconds.
\end{abstract}


\section{Introduction}


The Internet of Things (IoT) marketplace has experienced exponential increases over the past few years. According to \cite{Exp-growth}, IoT is reported to contribute anywhere between 4-11 percent of the total global GDP by 2025.
Unsurprisingly, IoT devices were first adopted in commercial settings such as smart building environments before 2017. After 2017, IoT started to explode into the smart home sector, where the smart home sector was valued at 79.13 billion in 2020. This growth is projected to continue to reach 319.25 billion dollars by 2026 ~\cite{mordor}.

However, the safe adoption of home IoT systems presents a unique set of challenges. 
One of these challenges is the heterogeneity of these devices. 
Due to the many different types of IoT devices developed by a variety of manufacturers, uniform management of these devices is challenging \cite{Hindawi, Type-Threat}. 
This is due to the explosive growth of demand for IoT home devices. 
Secondly, these devices are low on resources both in storage and computation, which creates difficulties in providing robust security solutions. 
By 2021, an anticipated $25\%$ of attacks on businesses would result from compromised IoT devices \cite{Type-Threat}. 
Thirdly, IoT systems generate massive amounts of data, evidenced by 6.2 Exabyte (EB) produced in 2018 with an anticipated $478\%$ increase through 2021, posing significant challenges to security-related data analysis \cite{Type-Threat, Hindawi}. 
These challenges require smart home systems to incorporate enhanced security visibility.

Traditional systems are inadequate due to the high volume of data and a large amount of connected devices in IoT deployments \cite{IoT_book, Sec_Smart_Home}. 
However, Software-Defined Networking (SDN) offers a promising solution. 
SDN centralizes network management and decouples the network and data planes.
By providing flow-based statistics, SDN offers immediate network visibility and simplifies intrusion detection \cite{Bizanis}. 
This poises SDN to offer an effective and efficient framework for handling heterogeneous networks while providing network visibility and control via flow monitoring and management of data plane.
Additionally, SDN-based systems can be augmented to develop managed and flexible solutions and intelligent decisions. 
By adopting machine learning techniques, these systems can offer promising solutions to enhance security in resource-constrained IoT home networks \cite{SDN/ML}. 
Firstly, by classifying devices, SDN policies can be crafted to specific device types such as cameras, hubs, switches, or triggers. This provides an insight into the home network and the ability to provide particular security policy and Quality of Service (QoS) for specific devices \cite{sivanathan2018classifying}. 
Secondly, machine learning based Distributed Denial of Service (DDoS) detection can  protect smart home IoT devices from denial of service that disrupts normal functionality \cite{9204688}. 
These attacks occur when a congregation of different devices sends a massive amount of packets to overwhelm the target device and disconnect it from the network. These attacks, whether resulting from compromised devices on the network or from external attacking devices, consume network resources and cause difficulties for smart home users \cite{btushirimpact}. 

For the home network use case, we identify several major design goals that are reflected in our contributions. 
First, the solution needs to be easy to use and cost-efficient for deployment. 
Building off of this, the solution must be optimized to take as few resources as possible and must be real-time. Next, the solution must have high accuracy in device classification and DDoS detection; particularly, false positives must be minimized in order to mitigate user interactions with the anomaly detection system. 
Finally, the system must be robust in order to handle different types of IoT devices and protect privileged devices from compromised IoT devices.
To these ends, our primary contributions are as follows.
\begin{itemize}
\item A novel SDN-based architecture is designed and implemented with a Faucet controller and Open vSwitch (OVS) using the GNS3 network simulator. The proposed architecture is deployable on a low-cost edge system, such as a Raspberry Pi, and uses VLANs to separate verified devices from unverified ones within a home network.
\item We propose a minimum set of stateless flow-based features to perform both device classification and DDoS detection on the smart home network. We achieve an average accuracy of $97\%$ in device classification and $98\%$ in DDoS detection.
\item We utilize non-cumulative statistics to increase classification and DDoS accuracy while reducing the controller and switch communication overhead on the SDN system.
\item Two real-world datasets from two different testbeds are used to validate the efficacy and reliability of the proposed features and machine learning models. 
Furthermore, the number of samples in the datasets are reduced to a minimum size to meet a $95\%$ accuracy threshold.
The accuracy and latency of K-Nearest-Neighbors (KNN), Random Forest (RF), and Support Vector Machine with linear kernel (LK-SVM) models are investigated in details with varying polling intervals. 
\end{itemize}

The rest of this paper is organized as follows. 
We overview the related work in Section \ref{ref:related_work}, discuss the proposed testing environment in Section \ref{ref:anomaly_detection}, and present the proposed secure home architecture scheme for device and DDoS classification in Section \ref{ref:propsed_archit}.
The classification and DDoS detection results are discussed in details in Section \ref{ref:results}, followed by a conclusion in Section \ref{ref:conclusion}.

\section{Related Work}\label{ref:related_work}
Although various IoT management approaches have been proposed \cite{6883583,8290717,Deep-Real,Bottleneck-Intrus, Lu_2021}, they struggle to adapt to the complex design requirements present in IoT networks. 
For example, Amaral et al. \cite{6883583} utilize Deep Packet Inspection (DPI) with distributed network nodes that act as `watchdogs'. These nodes act as lightweight network intrusion detection system (IDS) units that are dispersed throughout a network and buffer chunks of packets, which are then compared to the IDS access rules.
The `watchdogs' help to perform the heavy computational task of packet inspection in a distributed way, but the design still suffers from scalability issues due to the intense processing requirements for DPI. 
Similarly, researchers in \cite{8290717} propose RADAR, which can detect 90\% of DDoS attack data within 90 seconds regardless of the number of bots involved in the attack. However, RADAR can only detect 50\% of attack data within 60 seconds of operation, and struggles to create an IoT defense system with accurate real-time classifications. Again this is due to the heavy processing burden necessary for DPI.
Many other DPI systems \cite{Deep-Real,Bottleneck-Intrus, Lu_2021} show similar delay shortcomings or do not utilize resources efficiently.

SDN systems have been proposed to address the above challenges faced by DPI based solutions \cite{Hameed2018SDNBC,Bizanis_Kuipers,Salman}. Hameed et al. \cite{Hameed2018SDNBC} have identified many of these vulnerabilities and suggested that SDN can be a solution. Similarly, Bizanis et al. \cite{Bizanis_Kuipers} outline how SDN and Network Function Virtualization (NFV) technology can be combined with wireless IoT networks to enhance network management. Salman et al. \cite{Salman} have identified several privacy and security concerns in IoT and suggest that SDN is an optimal paradigm to manage heterogeneous networks and handle scalability complications.

Recently, the advantages of SDN are augmented by machine learning models, especially for enhancing security functionality. This is mostly realized by commercial vendors, while research efforts have been slow to leverage these technologies for security purposes \cite{SDN/Fog, SDN-Box}. Some of the first efforts relied on dedicated hardware middleboxes or inspection engineers. Liu et al. \cite{SDN-Box} presented a middlebox-guard that leverages SDN-based intelligent processing to minimize network latency and provide security functionalities. An SDN-based middlebox is placed into a network to handle all of the security processing. However, this system requires periphery hardware that raises cost for the end user. Similarly, Han et al. \cite{han2018overwatch} propose OverWatch, a program that constantly monitors every flow to classify attacks. However, the flow inspection technique noticeably increases the overhead of the system. Both of these works add either additional hardware or extra computational requirements that make these systems difficult or too expensive to deploy for home networks.

Some of the most recent works focus on leveraging both SDN and machine learning without requiring periphery dedicated hardware. Hamza et al.\cite{hamza2019detecting} perform enhanced anomaly detection by translation of MUD profiles for IoT gateways. They use both stateful and stateless network statistics to detect anomalous traffic patterns in a MUD-compliant network. Specifically, the system mitigates benign and volumetric attacks with very high accuracy. Furthermore, work from Doshi et al. ~\cite{Doshi} and Yang et al. ~\cite{Yang} show an accuracy of $99.8\%$ for DDoS detection using a customized testbed and the KD99 dataset, respectively. This combination of machine learning and SDN has also been applied for device classification. Similarly work from Owusu et al. ~\cite{Owusu} has subsecond inference speeds for device classification in IoT environments with peak accuracy of $92.7\%$. Reza et al. ~\cite{Reza} and Xu et al. ~\cite{Xu} also propose methods for low-latency device classification, which achieve accuracy as high as $97.6\%$.  
However, recent works do have certain drawbacks. Specifically, these works do not fully consider the spacing and latency requirements for a gateway at the edge. The SDN controller is resource constrained and thus must be able to operate in edge environments. Furthermore, many of the machine learning models require a significant number of features. This requires a high amount of storage resources and creates additional dependencies that limit deployment. Moreover, most of the existing works only use one dataset which may lead to models overfitting the singular dataset, whereas training with multiple datasets can make models more robust to dynamic changes that are seen in IoT environments. Finally, to our knowledge, no previous work has proposed an SDN-based solution that combines both classification and DDoS detection into a single model. Our work seeks to combine both of these systems to a single and unified machine learning approach which saves space and time compared to separate models.

It is important to note that some of the features proposed in this paper, such as ICMP, TCP, and UDP distribution, packet size and packet count, have been adopted in prior works ~\cite{hamza2019detecting,DDoS_DPI_2, DDoS_Flow} for DDoS detection. However, even with these features, existing literature typically requires more than fifteen features to successfully perform DDoS detection. In this work, we propose a unique feature as IP diversity ratio, which is combined with ICMP, TCP, and UDP distribution, packet size and count, to form a minimal set of features. More importantly, this minimum set of features are used for not only DDoS detection, but also device classification and reducing processing overhead.
In summary, we propose a secure smart home architecture by adopting SDN and machine learning models, which performs both device classification and DDoS detection with high accuracy. 
The stateless and flow-based features proposed in this paper are efficient to extract, and the low-overhead of this solution distinguishes this paper from the current literature by considering memory constraint and reducing computational complexity.

 \begin{figure}[!t]
 \centering
    \includegraphics[width=0.8\linewidth]{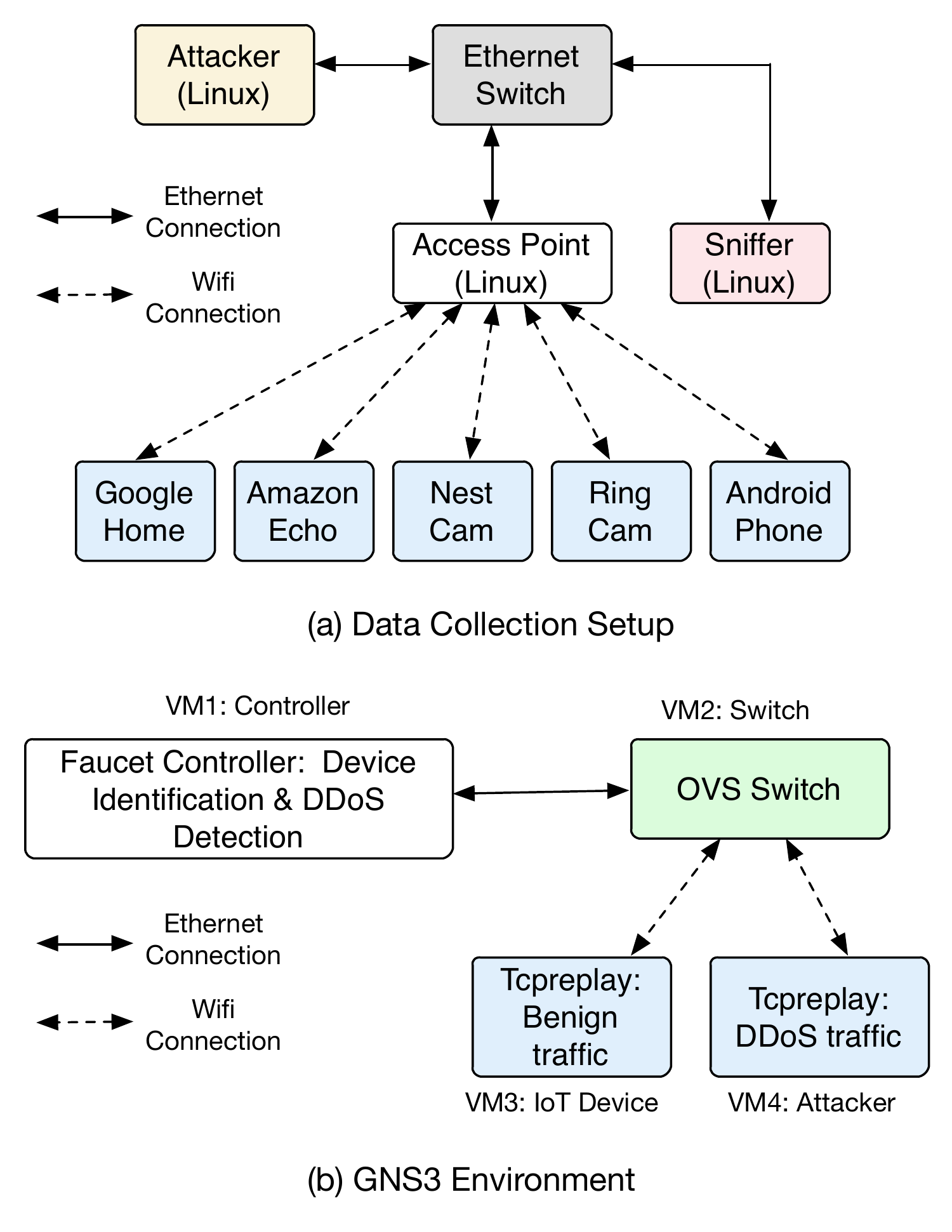}
    \vspace{-3mm}
    \caption{(a): Testbed setup used for benign and DDoS dataset collection. (b) A virtualization-based system setup using the GNS3 simulator.}
    \label{data_col}
    \vspace{-2mm}
\end{figure}

\section{Proposed Testing Environment}
\label{ref:anomaly_detection}
This section discusses the testbed setup, the GNS3 virtual environment's overview, and the proposed features for IoT device classification and DDoS attack detection.

\subsection{Smart Home Network} 
\label{traf_col}

We setup a smart home network with consumer IoT devices to collect benign and DDoS attack data, as shown in Figure \ref{data_col} (a). For benign data collection, we configure two Linux machines. One machine acts as a WiFi access point (AP), and another machine acts as a sniffer to log the packets in the pcap (packet capture) file. The proposed testbed includes the following IoT devices: Google Home, Amazon Echo, Nest Camera, Ring Camera and an Android phone. We install  {\fontfamily{pcr}\selectfont hostapd} on the AP machine to enable AP functionalities. Also,  {\fontfamily{pcr}\selectfont tshark} is installed on the sniffer for packet collection. {\fontfamily{pcr}\selectfont tshark} is a command line tool, which helps to reduce the GUI overhead on Linux machines. 

The IoT devices are connected to the WiFi AP. To collect IoT devices' benign traffic, we interact with each IoT device for 24 hours and record the pcap files. These interactions are launched based on real activities that would occur during regular device usage, including streaming videos from Nest and Ring Camera, playing songs, and asking questions from Amazon Echo and Google Home.

To collect DDoS attack traffic, the following three typical DDoS attacks are generated: ICMP, TCP-SYN, and UDP. 
A Linux machine is used as the attacker. 
Also, the DDoS attacks are launched by the {\fontfamily{pcr}\selectfont hping3} utility, which helps to dynamically change the IP source address, IP destination address, attack type, and attack duration. For each type of DDoS attacks, the attacker targets each IoT device for 10 minutes while the sniffer records the pcap files of the attack traffic. This process produces a dataset including 3,066,585 packets, composed of 1,264,392 malicious packets and 1,802,193 benign packets.

To confirm the proposed features' robustness, in addition to our lab (SIOTLAB) dataset, we also use the UNSW dataset ~\cite{hamza2019detecting}, consisting of 802,580 benign packets. UNSW dataset comprises ten smart home devices including Wemo Motion Sensor and Power Switch, Samsung and Netatmo Camera, TP Link Smart Plug, Hue Bulb, Amazon Echo, Chromecast, iHome, and LiFX ligthbulb. 
Next, the DDoS packets are consolidated with SIOTLAB and UNSW benign dataset for the DDoS detection. 
Also, we combine SIOTLAB's dataset and UNSW's dataset to ensure that the machine learning models are not biased.

\subsection{GNS3 Environment} \label{GNS3_env}
We build a virtual SDN-based smart home environment using the GNS3 network simulator.
Figure \ref{data_col} (b) presents this environment, which includes four virtual machines (VMs). 
The first VM acts as the Faucet SDN controller. 
The second VM acts as an OVS switch, which is an open-source OpenFlow switch. We choose the OVS switch as it is well suited to function as a virtual switch in virtualized environments. 
The third VM operates as an IoT device, and the fourth VM serves as an attacker.

\begin{figure}[!t]
    \includegraphics[width=1\linewidth]{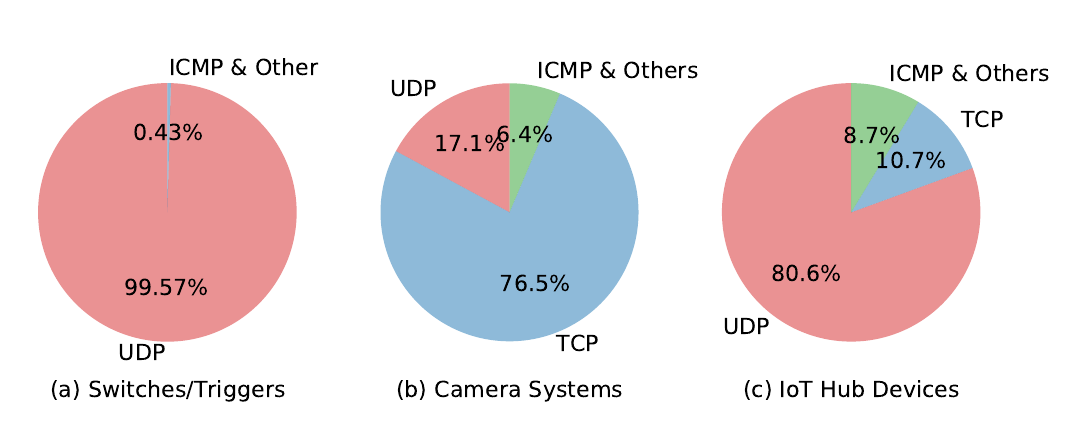}
    \caption{The distinct difference between the distribution of TCP, UDP, ICMP, and other flows among device types enables us to successfully classify devices.}
    \label{pie_chart}
    \vspace{-2mm}
\end{figure}

\begin{figure*}
    \includegraphics[width=7.1in]{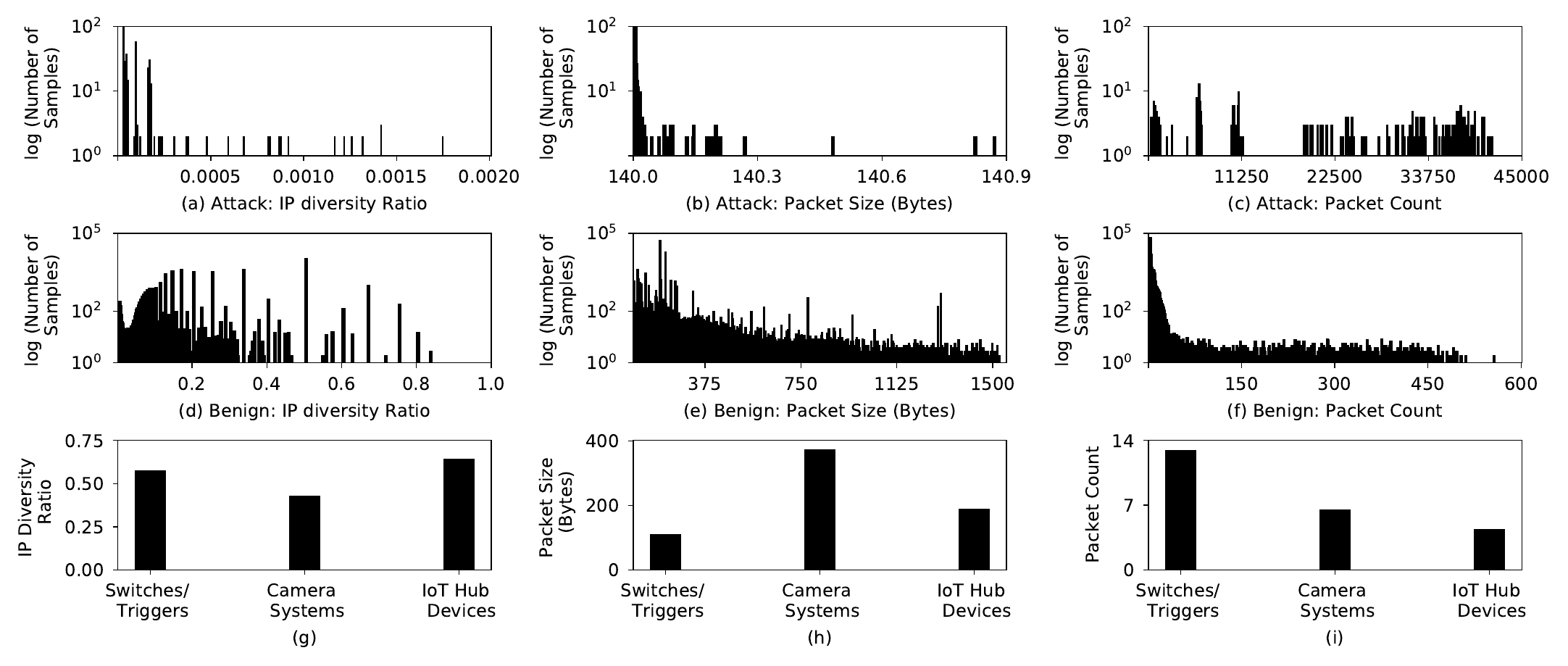}
     \caption{The higher shift in IP diversity ratio, packet size, and packet count allows us to classify devices and identify DDoS attacks with a threshold accuracy of $95\%$. All of these values are obtained from flow data with an optimized polling interval of 24 seconds. The benign samples are collected for 24 hours while the attack samples are collected over 10-minutes time frames.} 
    \vspace{-2mm}
    \label{ddos_ratio}
    \vspace{-2mm}
\end{figure*}
To evaluate the classifier's performance, we use {\fontfamily{pcr}\selectfont tcpreplay} to playback the benign and DDoS packets. {\fontfamily{pcr}\selectfont tcpreplay} is a  Linux tool that helps in replaying previously captured network traffic. 
Besides, it is commonly used to simulate attacks to test intrusion detection systems. To determine each dataset's performance, namely, SIOTLAB, UNSW, and the integration of SIOTLAB with UNSW, we replay packets belonging to each dataset one by one and test the classification and DDoS detection of the proposed solution.

In traditional SDN networks, the controller polls all the network switches periodically to collect flow statistics. 
In response, switches send cumulative statistics of sent and received packets to the controller. To generate accurate flow statistics, the controller polls at lower intervals which increases the network congestion between the switch and the controller. This highlights an important tradeoff between the accuracy of the statistics gathered from the switch and the network management overhead ~\cite{payless, poll-acc}. 
However, in this work, we seek to mitigate this tradeoff by using non-cumulative statistics instead of cumulative statistics. 
This means that every poll from the controller to the switch resets the flow gauges on the controller. 
It is important to note that this functionality is not supported on many SDN controllers; therefore, it is accomplished by reading the latest polling interval's flow details from the switch and subtracting the previous poll value. 
This allows each non-cumulative statistic to be treated independently; hence, it increases visibility into the new interval by not including the events that happened before this period.
To identify the optimized non-cumulative polling length, we train the classifier for varying polling intervals from 1 to 120 seconds for each dataset.

\subsection{Proposed Stateless Flow-based Features}
We explore stateless flow-based features and analyze their suitability for IoT device classification and DDoS detection. 
These features can be generated without splitting the incoming traffic streams based on IP sources, making them flow-based features. These features are also stateless because only easily accessible header information is used and the packets are never decapsulated. Furthermore, these proposed features are adopted for both device classification and DDoS detection, making it efficient for feature extraction. 

All of the IoT home devices present in the SIOTLAB and UNSW datasets are listed in Table \ref{device_type}. 
Please note that even though we listed switches and triggers as one device category, they are distinct. Switches are the devices that help turn on/off the connected appliances, such as the TP-Link Smart Plugs, while triggers are the devices that activate on an event, for example, WeMo Motion Sensors. 
The average distribution of TCP, ICMP, UDP, and other protocols reveals three important patterns. Firstly, Figure \ref{pie_chart} (a) shows that the percentage of UDP traffic with switches/triggers is $99.57\%$. 
This is because switches/triggers require minimal delay and exchange very small amounts of data. 
Therefore, they can perform application-layer-controlled retransmissions to ensure reliability \cite{TCP}.
The switches/triggers also generate the burst flows, which inflates the average number of packets per non-cumulative statistic as seen in Figure \ref{ddos_ratio} (i).  

Secondly, Figure \ref{pie_chart} (b) shows camera systems have adopted the highest TCP percentages in the dataset. This is due to the increase in popularity of TCP, as modern camera systems are integrated into front-end connection-oriented application platforms \cite{TCP}. 
Also, camera systems employ large payload sizes for images to be displayed on front-end applications as seen in Figure \ref{ddos_ratio} (h). 
Cameras exchange an average of 6.9 packets per non-cumulative statistic, which is more than that of hubs and less than that of switches/triggers as seen in Figure \ref{ddos_ratio} (i).
This is due to cameras requiring to transmit large amounts of image data while also using large payload sizes in the packets. This keeps the average packet count below those of switches/triggers, but also greater than IoT hubs. The camera systems also have a low IP diversity ratio as seen in Figure \ref{ddos_ratio} (g); therefore, they mainly interact with one to two other devices in any given non-cumulative statistic.

Thirdly, Figure \ref{pie_chart} (c) shows that IoT hub devices generate UDP packets because they are mainly real-time voice assistants which require minimal latency, but at the same time they show a higher TCP percentage because they typically communicate with external services to respond to user queries and requests \cite{TCP}. These devices typically interact with a variety of unique IP destination addresses and send high number of packets. This causes the IP diversity ratio to average at around 0.6 as seen in Figure \ref{ddos_ratio} (g). This means that hub devices talk to many different users at once and spread out their traffic to respond to these destination IP addresses. This is because hubs have to communicate with users and backend servers to respond to user requests. Therefore, their communication typically demonstrates bursty flows with small payload sizes that are destined to a wide variety of IP addresses. 
Finally, these systems tend to have the idlest time, with an average of 4.8 packets per non-cumulative statistic as seen in Figure \ref{ddos_ratio} (i).

\begin{table}[!tb] 
 \centering
\def\arraystretch{1.2}
    \caption{IoT Device Category and Type: SIOTLAB and UNSW dataset}
        \begin{tabular}{|p{2.5cm}|p{4.5cm}|}
        \hline
                  Device Category  & Device \\ \hline         \hline
                  Switches/Triggers    &  WeMo Motion  Sensor and Power Switch, TP Link Smart Plug, Chromecast  \\ \hline
                  Camera Systems   &   Ring Camera, Nest Camera, Samsung Camera, Netatmo Camera \\ \hline
                  IoT Hub Devices  &  Amazon Echo, LiFX, Hue Bulb, iHome, Google Home \\ \hline
           
\end{tabular}
\label{device_type}
\end{table}

When looking at the differences between attack and benign data, important differences emerge. Firstly, DDoS attacks typically target one victim device. This drastically drives down the IP diversity ratio. This is is highlighted in Figure \ref{ddos_ratio} (a) and (d). 
The IP diversity ratio during attacking scenarios is underneath 0.002 which is significantly lower than in benign flows. Secondly, devices seeking to perform a DDoS attack will attempt to flood a system; therefore, packet size is minimized in order to get the highest throughput for packets highlighted in Figure \ref{ddos_ratio} (b) and (e). 
Finally, the packet count per non-cumulative statistic is also significantly higher during an attack period versus a benign period as seen in Figure \ref{ddos_ratio} (c) and (f). This allows these features to perform both classification and DDoS detection. These features are summarized below:

\begin{itemize}
    \item \textbf{Protocol percentage:} For each polling interval, the percentage of ICMP, TCP, and UDP packets is calculated. Figure \ref{pie_chart} shows the traffic generated by the IoT devices exhibit varying protocol distributions that help to classify IoT devices. 
    Summarily, this distribution changes whenever an attack occurs. For example, TCP-SYN attacks cause the percentage of TCP packets to increase from a device's normal distribution of features. 
    \item \textbf{IP diversity ratio:} This metric is calculated as the number of unique IP addresses divided by the total packets sent by a device in a polling interval. Typical IoT devices usually communicate with a limited number of devices and servers on their network, while victim devices typically receive attack packets from much more diverse attacking devices. This causes the attacking flows to have a significantly lower IP diversity ratio compared to benign devices. Figures \ref{ddos_ratio} (a) and (d) highlight this effect. 
    \item \textbf{Packet count and size:} These features are based on the observation that the distribution of packet size differs significantly among IoT devices and among benign and DDoS attacks. The attacker keeps the size of the attack packet as small as possible to avoid extra overhead, and the payload size typically small and has minimal variance in DDoS attacks \cite{Doshi}. This is due to flooding DDoS attacks requiring high transmission rates. Furthermore, DDoS attacks typically have high packet counts per non-cumulative statistic. These observations are shown in Figures \ref{ddos_ratio} (b), (c), (e), and (f).  
   
\end{itemize}

\section{Proposed Secure Home Architecture} \label{ref:propsed_archit}
The proposed architecture operates in two phases: 
in phase I, the SDN controller places a device in the unverified VLAN with lower QoS. 
Later, the OVS switch generates the flow data. 
The Faucet controller collects this flow data by polling the OVS switch. 
Later, the Faucet controller extracts the features on a per-device basis (using their IP addresses) and feeds the features to the classifier.
The classifier then predicts whether each device is malicious or benign along with the device category. Suppose a device falls into one of the categories as shown in Table \ref{device_type}. In that case, the controller moves it to the verified VLAN; else, it is flagged and remains in the unverified VLAN, and the second phase is initiated. During phase II, for the flagged devices, the controller runs the classifier, and if an attack is detected, the device is removed from the network, blocking the attack.

\begin{figure*}
    \includegraphics[width=7.1in]{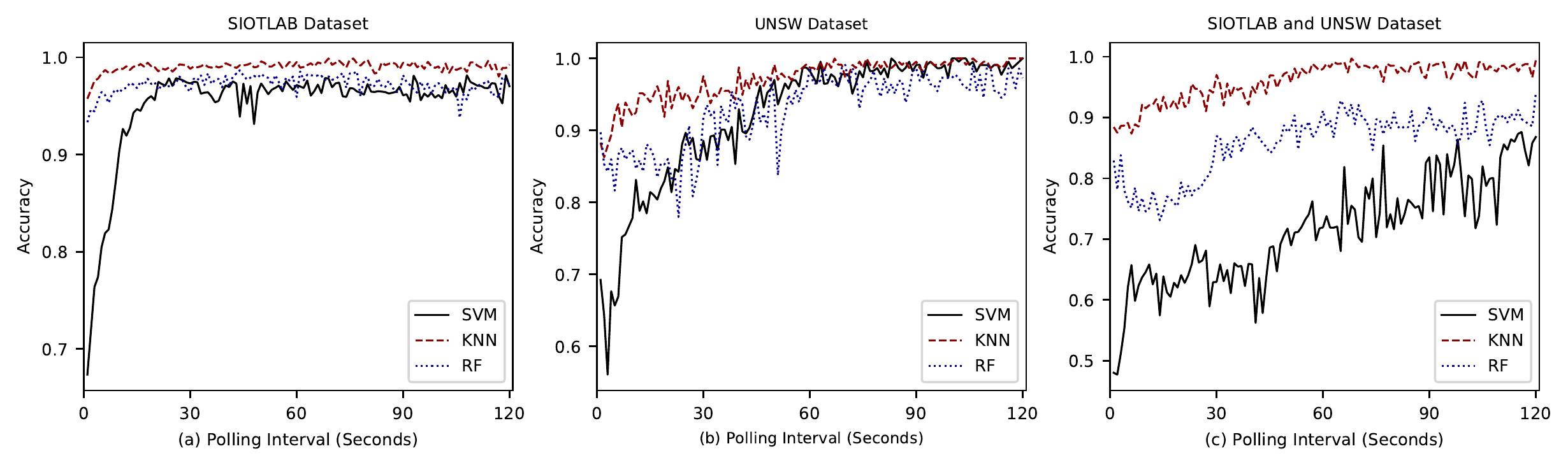}
    \caption{Accuracy of KNN, RF, and LK-SVM for (a) SIOTLAB dataset, (b) UNSW's dataset, and (c) the combination of both datasets. 
    All models demonstrate accuracy improvement versus increasing polling interval. }
    \label{acc_sampling_window}
\end{figure*}

\section{Result and Discussion}
\label{ref:results}
We test three machine learning models for IoT device classifications and DDoS detection. 
\begin{itemize}
    \item K-Nearest-Neighbors (KNN)
    \item  Support Vector Machine with linear kernel (LK-SVM)
    \item Random Forest using Gini impurities (RF)
\end{itemize}
{These machine learning models are chosen because of their widespread usage in the literature for IoT device classification and DDoS detection \cite{DDoS_Flow,Hameed2018SDNBC, aung2017detection,hamza2019detecting}}.
We use these machine learning models implemented in the Scikit-learn Python library \cite{pedregosa2011scikit}. 
Since our major focus is to evaluate the effectiveness of the proposed feature set, all the hyper-parameters are the default unless otherwise stated.
We split each dataset into $75\%$ training data and $25\%$ testing data. 
Furthermore, each class type is balanced in order to ensure strong validation results and to mitigate bias in the machine learning models.

\begin{figure}[!t]
    \includegraphics[width=.475\textwidth]{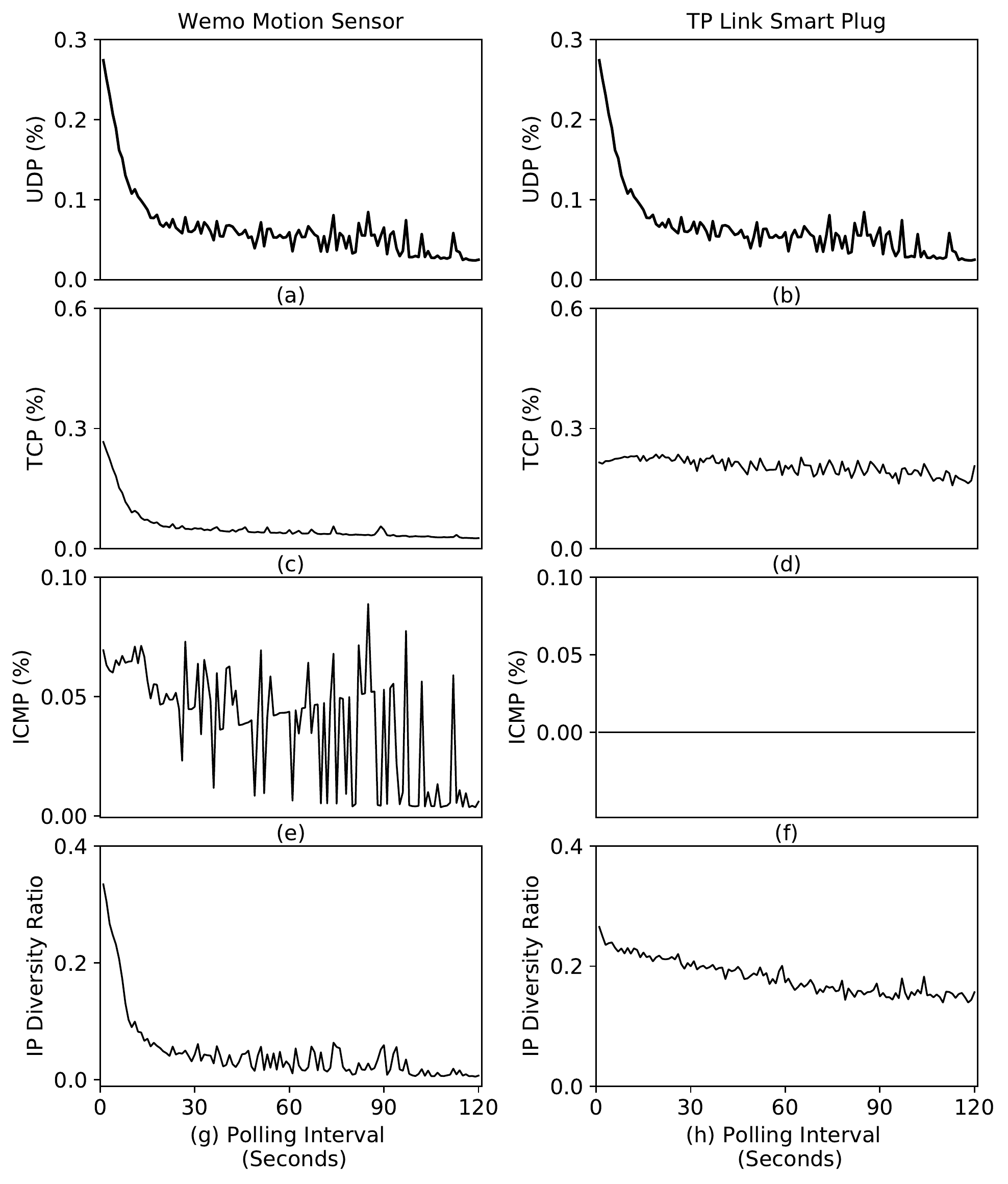}
    \caption{(a)-(f): Normalized standard deviation of the percentages of packets belonging to various protocol types. 
    (g)-(h): Normalized standard deviation of the diversity of the destination IP address of packets.
    The left column present the results for Wemo Motion Sensor, and the right column represents the results for TP Link Smart Plug.
    }
    \label{std_unsw}
\end{figure}

To determine the association between polling intervals and the accuracy of machine learning models, we train and test them against various non-cumulative statistics gathered at different polling intervals. 
Figure \ref{acc_sampling_window} shows that an increase in polling interval results in a higher accuracy for all three models (KNN, LK-SVM, and RF). 
The is because compared to a small polling interval, more non-cumulative statistics are collected during a long polling interval, and this results in lower variations of the features.
This is seen in Figure \ref{std_unsw} where the standard deviation of the features remains below 0.1 or drops drastically as the polling interval is increased. 
This allows the features to stabilize without utilizing historic data as long as the polling interval is sufficiently large.
Employing longer polling interval also reduces network management overhead.
However, it is important to note that long polling intervals decrease the responsiveness of the SDN controller. 
For example, if a polling interval is 20 seconds, the controller can only receive new data every 20 seconds and process it before making a decision. 
In contrast, a 50-second polling interval takes a minimum of 50 seconds to detect an attack. 
Consequently, a SDN controller must mitigate attacks in an adversarial environment before substantial damage to home devices can occur. Therefore, an optimal polling interval must be selected to establish a tradeoff between responsiveness, accuracy, and overhead.  


When selecting the optimal polling interval, 24 seconds is chosen for two primary reasons. 
Firstly, among KNN, LK-SVM and RF, KNN can achieve accuracy higher than $95\%$ for polling intervals longer than or equal to 24 seconds.
Secondly, our prior work \cite{9204688} has identified that the minimum attack duration to disconnect an IoT device from the AP is 120 seconds.
Hence, using 24 seconds as the polling interval leaves the model with enough time to block or remove devices attacking the smart home network before a device is disconnected \cite{9204688}. 
This balances the needs for system responsiveness, high accuracy, and low overhead. 

\begin{table}[!tb] 
 \centering
\def\arraystretch{1.2}
    \caption{Traffic Classification Results (SIOTLAB Dataset)}
        \begin{tabular} {|p{2.1cm}|p{1.5cm}|p{1.5cm}|p{1.5cm}|p{1.5cm}|}
        \hline
        Class          &  KNN & LK-SVM  & RF  \\ \hline  \hline
        Switches/Triggers & $99.2\%$ & $87.0\%$ & $98.6\%$\\ \hline  
        Camera Systems  & $98.1\%$& $93.3\%$& $96.5\%$\\ \hline    
        IoT Hub Devices & $99.3\%$& $92.1\%$ & $98.9\%$\\ \hline    
        DDoS Attacks & $99.9\%$& $94.9\%$& $99.1\%$ \\ \hline
\end{tabular}
\label{Table-SIOT}
\end{table}

\begin{table}[!tb] 
 \centering
\def\arraystretch{1.2}
    \caption{Traffic Classification Results (UNSW)}
        \begin{tabular} {|p{2.1cm}|p{1.5cm}|p{1.5cm}|p{1.5cm}|p{1.5cm}|}
        \hline
        Class          &  KNN & LK-SVM  & RF  \\ \hline  \hline
        Switches/Triggers & $96.7\%$ & $78.2\%$ & $77.3\%$\\ \hline  
        Camera Systems  & $97.1\%$& $72.8\%$& $81.2\%$\\ \hline    
        IoT Hub Devices & $96.0\%$& $69.3\%$ & $78.6\%$\\ \hline    
        DDoS Attacks & $97.8\%$& $79.5\%$& $80.8\%$ \\ \hline
\end{tabular}
\label{Table-UNSW}
\end{table}

\begin{table}[!tb] 
 \centering
\def\arraystretch{1.2}
    \caption{Traffic Classification Results (SIOTLAB and UNSW dataset)}
        \begin{tabular} {|p{2.1cm}|p{1.5cm}|p{1.5cm}|p{1.5cm}|p{1.5cm}|}
        \hline
        Class          &  KNN & LK-SVM  & RF  \\ \hline  \hline
        Switches/Triggers & $94\%$ & $65.2\%$ & $75.0\%$\\ \hline  
        Camera Systems  & $93.1\%$& $67.8\%$& $73.9.2\%$\\ \hline    
        IoT Hub Devices & $96.2\%$& $59.3\%$ & $71.6\%$\\ \hline    
        DDoS Attacks & $99.2\%$& $73.2\%$& $80.8\%$ \\ \hline
\end{tabular}
\label{Table-both}
\end{table}

When evaluating the datasets, KNN performs the best with $99.8\%$ average accuracy in the SIOTLAB dataset.
Furthermore, KNN also performs the best with $97.9\%$ accuracy for the UNSW dataset. 
Finally, for the combined dataset, KNN accuracy remains high at $96.5\%$; however, the average accuracy of LK-SVM and RF drops significantly highlighted in Figure \ref{acc_sampling_window} (a)-(c). To identify the reason for the varying performance of models, we analyzed the data and reached the following conclusions. 
First, the data is  tightly clustered among samples belonging to the same device and category. 
This helps KNN to pull the $k$ nearest neighbors accurately. 
Second, the data is not axis-aligned; thus, RF, an ensemble of axis-aligned decision trees \cite{tomita2020sparse} does not perform well. Finally, LK-SVM does not achieve high enough accuracy due to the difficulty of cutting decision regions into linear hyperplanes. 
This is because clusters can create dimensional spaces that are significantly more difficult to cut into linear regions; hence, KNN proves to be a suitable model for device classification and DDoS detection.

This proposed scheme's accuracy with the KNN model makes it a strong competitor with existing works. 
For examples, Hamza et al. ~\cite{hamza2019detecting} achieved peak accuracy at $97.5\%$ while this work combines their dataset (UNSW) and our own (SIOTLAB) to achieve an accuracy of $98\%$. 
Furthermore, their work uses 20 total features while the proposed system uses six features.
\begin{figure}
\centering
    \includegraphics[width=.8\linewidth]{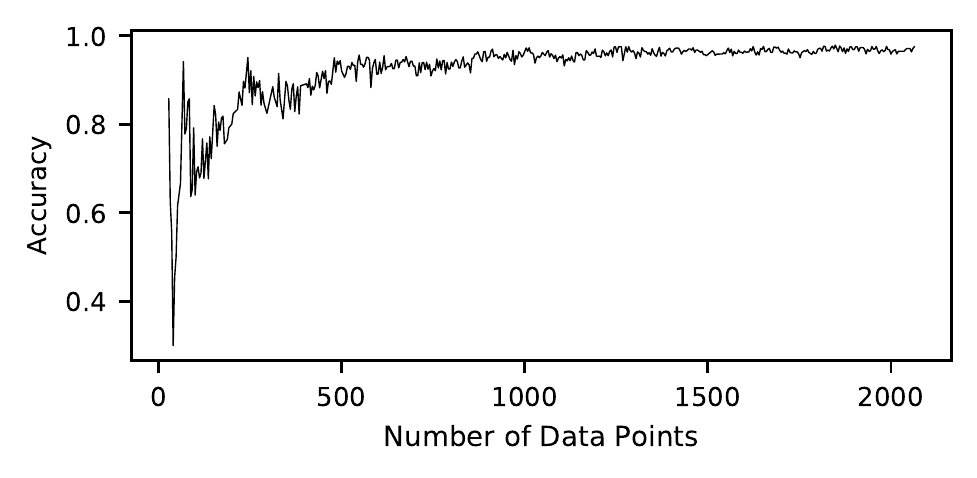}
    \caption{For the combined dataset, using KNN shows 95\% accuracy for 1820 data points, meeting the low memory requirement.}
    \label{fig:KNN_DDoS}
\end{figure}

\begin{figure}
\centering
    \includegraphics[width=0.8\linewidth]{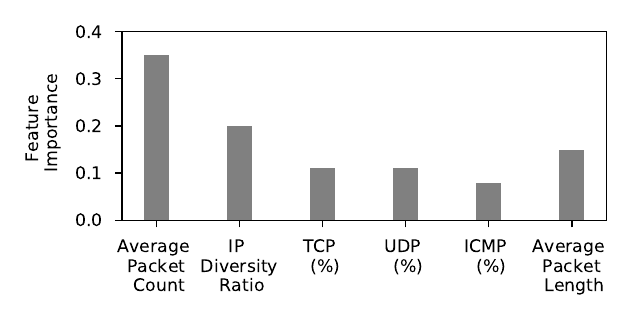}
 \caption{The relative importance of all six features when applying KNN on the combined dataset with a 24-second optimal polling interval. }
    \label{feat_importance}
\end{figure}

\begin{table*}[t] 
\centering
\def\arraystretch{1.5}
    \caption{Comparison with Existing Work}
        \begin{tabular} {|p{2.1cm}|p{1.5cm}|p{2cm}|p{3cm}|p{1cm}|p{1cm}|p{1cm}|p{1cm}|p{1.5cm}}
        \hline
        Existing Work          & Maximum Accuracy & Detection Type  & Dataset & Storage & Latency & Feature Count\\ \hline  \hline
        
        Owusu et al. ~\cite{Owusu} & $92.7\%$ & Classification & Tor dataset & No & No & 6\\ \hline  
        Reza et al. ~\cite{Reza} & $97.6\%$ & Classification & Custom dataset & No & No & 13\\ \hline  
        Xu et al. ~\cite{Xu} & $87.8\%$ & Classification & Custom dataset & No & No & 21\\ \hline  
        Hamza et al. ~\cite{hamza2019detecting} & $97.5\%$ & DDoS & UNSW & Yes & Yes & 20\\ \hline  
        Doshi et al. ~\cite{Doshi} & $99.8\%$ & DDoS & Custom dataset & No & No & 11\\ \hline  
        Yang et al. ~\cite{Yang} & $99.8\%$ & DDoS & KD99 & No & No & 9\\ \hline

        \textbf{Proposed solution} & $\textbf{99.9}\%$ & \textbf{Classification and DDoS} & \textbf{Custom (SIOTLAB) and UNSW Dataset} & \textbf{Yes} & \textbf{Yes} & \textbf{6} \\ \hline 
\end{tabular}
\label{Comparison}
\end{table*}

The final consideration for smart home deployment is the memory requirements. 
Smart home IoT gateways do not have high amounts of physical memory; therefore, the dataset must be reduced as much as possible. Figure \ref{fig:KNN_DDoS} shows the variation of KNN accuracy with training dataset size for combined dataset and 24-second polling interval. 
A dropout from each class of the data is performed in order to minimize the dataset size. 
Also, the non-cumulative statistics that only include signaling information such as DNS and NTP are removed. 
The total dataset is reduced to 1820 data points before crossing the $95\%$ threshold.

To determine the KNN model's latency for the 24-second non-cumulative statistics, we test it with the training dataset that includes 1820 data points. This is important because the KNN model scans its historical training data before each inference. Thus, KNN often has a slow inference speed. To measure the KNN's average latency, we run approximately a hundred trials of device classification and DDoS detection on a machine with a 1.4 GHz i5 processor and 4 GB of RAM. These trials yield an average latency of 1.18 milliseconds (ms).
This optimized storage of training data and low latency allows for more accessible lightweight edge deployments. 

The relative feature importance after minimization of combined dataset and KNN is plotted in Figure \ref{feat_importance}. This feature importance is calculated by performing a permutation of the features and randomly dropping subsets of the feature set. This results in an accuracy drop that is converted into a percentage for the relative impact of each feature on the classification for any model. This is accomplished by using the SciKit Learn feature permutation function from  ~\cite{pedregosa2011scikit}. As shown above, the IP Diversity Ratio contributes approximately $20\%$ to the device classification and DDoS detection efficacy. This suggests that this feature is a valuable addition for both device classification and DDoS detection.

Table \ref{Comparison} shows various metrics used to compare the proposed solution with existing solutions. 
The proposed solution yields a high accuracy while performing both device classification and DDoS detection. We also measure latency and storage requirements for realistic edge deployment. Further, the results are validated by using data from two different testbeds. The proposed solution uses a limited feature set, specifically only six features. This makes the training data for our system significantly smaller than that of other works. The optimal KNN model in this work takes up 541.282 KB. The work from Hamza et al. ~\cite{hamza2019detecting} had a model size of approximately 14.2 MB. However, it is fair to note that their model did rely on optimizations routinely used in scikit learn. Nonetheless, the proposed model outperforms other models by reducing the trained model size from 14.2 MB (in \cite{hamza2019detecting}) to 541.282 KB. 


When comparing latency, it is essential to note that different machines will have heterogenous hardware components that can allow for different latencies. Moreover, the RAM and CPU clocking speed can drastically affect latency time. Therefore, we just present the hardware setup used in different works and their corresponding latency. In particular, our experiments are conducted on a machine with a 1.4 GHz Intel Core i5 processor and 4 GB of RAM, and achieved an average latency of 1.18 ms. Hamza et al.'s ~\cite{hamza2019detecting} achieved 13 ms latency time on an Intel Core CPU 3.1 GHz laptop with 16 GB of RAM. 
Finally, our system performs both device classification and DDoS detection, which extends prior solutions that only focused on either device classification or DDoS detection.

\section{Conclusion}\label{ref:conclusion}
In this work, we propose a SDN-based security architecture for smart homes. 
In particular, we perform IoT device classification and DDoS detection using a minimal set of features, including the percentage of ICMP, TCP, and UDP flows, packet size, packet count, and IP diversity ratio. 
We also evaluate the following machine learning models: KNN, LK-SVM and RF. 
The proposed solution meets the memory, latency, and security requirements by reducing the training dataset size, selecting an appropriate machine learning model, providing stateless and easy-to-calculate features, and increasing the polling interval to reduce overhead on SDN controller. 
We observe that for the proposed feature set and longer polling intervals (i.e., more non-cumulative statistics), KNN performs better than LK-SVM and RF. 
Also, for 1820 data points and a 24-second optimal polling interval, KNN achieves greater than $95\%$ accuracy with an average latency of 1.18 milliseconds.

\bibliography{references.bib}
\bibliographystyle{IEEEtran}
\end{document}